\documentclass[10pt,twocolumn,english,aps,manuscript,final,5p,times,prl]{revtex4}
\usepackage[T1]{fontenc}
\usepackage[latin9]{inputenc}
\usepackage{amsmath}
\usepackage{amssymb}
\usepackage{graphicx}
\usepackage{esint}

\makeatletter
\@ifundefined{textcolor}{}
{%
 \definecolor{BLACK}{gray}{0}
 \definecolor{WHITE}{gray}{1}
 \definecolor{RED}{rgb}{1,0,0}
 \definecolor{GREEN}{rgb}{0,1,0}
 \definecolor{BLUE}{rgb}{0,0,1}
 \definecolor{CYAN}{cmyk}{1,0,0,0}
 \definecolor{MAGENTA}{cmyk}{0,1,0,0}
 \definecolor{YELLOW}{cmyk}{0,0,1,0}
}

\makeatother

\usepackage{babel}
\begin{document}

\title{Rise of pairwise thermal entanglement for an alternating Ising
and Heisenberg spin chain in an arbitrarily oriented magnetic field }

\author{M. Rojas, S. M. de Souza and Onofre Rojas}

\affiliation{Departamento de Ci\^encias Exatas, Universidade Federal de Lavras.
CP 3037, 37200-000, Lavras -MG, Brazil.}
\begin{abstract}
Typically two particles (spins) could be maximally entangled at zero temperature, and for a certain temperature the phenomenon of entanglement vanishes at the threshold temperature. For the Heisenberg coupled model or even the Ising model with a transverse magnetic field, one can observe some rise of entanglement even for a disentangled region at zero temperature.  So we can understand this emergence of entanglement at finite temperature as being due to the mixing of some maximally entangled states with some other untangled states. Here, we present a simple one-dimensional Ising model with alternating Ising and Heisenberg spins in an arbitrarily oriented magnetic field, which can be mapped onto the classical Ising model with a magnetic field. This model does not show any evidence of entanglement at  zero temperature, but surprisingly at finite temperature rise a pairwise thermal entanglement  between two untangled spins at zero temperature, when an arbitrarily oriented magnetic field is applied. This effect is a purely magnetic field ,and the temperature dependence, as soon as the temperature increases, causes a small increase in concurrence achieving its maximum at around 0.1.  Even for  long-range entanglement, a weak concurrence still survives. There are also some real materials that could serve as candidates that
would exhibit this effect, such as Dy(NO$_{3}$)(DMSO)$_{2}$Cu(opba)(DMSO)$_{2}$\cite{hagiwara-strecka}.
\end{abstract}


\maketitle
Condensed matter researchers have long investigated the correlation
function between parts of composite systems, thus it is relevant
to study the quantum part of these correlations, which is called entanglement.
Quantum entanglement is one of the most fascinating features of quantum
theory due to its nonlocal property. Therefore, quantum entanglement
has been the subject of many research in recent years as a potential
resource for information processing and quantum computing. 

Quantum entanglement , with  its applications  to quantum  phase transitions of strongly correlated spin systems and its  experimental implementation in optical lattices, was considered, in particular, for one-dimensional systems.
Diverging entanglement length without quantum phase transition was found in localizable entanglement (LE) for VBSs,
since the correlation length remains finite\cite{vestraete}. This is a rather  remarkable result regarding the
entanglement properties of VBS quantum spin ground states. A  theory for localizable entanglement was developed based on
matrix product states coming from DMRG method and applied to VBS states\cite{popp}. In reference \cite{garcia}, an  experimental implementation was proposed for VBSs of  spin-1 Heisenberg Hamiltonians and ladders, and a method was proposed to directly measure quantum observables that are not accessible in standard materials in condensed matter.

Recently,  special attention has been focused on synthesizing
Ising-Heisenberg chains, which are well represented as spin systems
composed of Ising (classical) and Heisenberg (quantum) spins\cite{valverde08,ohanyan09,antonosyan09,canova09,rojas-ohanyan,ana-roj,strecka-jasc05,vanden10,sahoo12}.  Ising-Heisenberg chains plays an important role in providing evidence
for several novel and unexpected quantum states \cite{valverde08,ohanyan09,antonosyan09,canova09,rojas-ohanyan},
such as thermal entanglement, fractional magnetization plateaus in
the low-temperature magnetization process \cite{antonosyan09,canova09,rojas-ohanyan}, 
and so on. Thus, real materials with experimental realizations of
 Ising-Heisenberg chains can be represented by the aforementioned
theoretical findings, such as the magnetic behavior of a tetramer Ising-Heisenberg
bond-alternating chain as a polymeric model Cu(3-Clpy)$_{2}$(N$_{3}$)$_{2}$
\cite{strecka-jasc05}. Another magnetic polymer is the dysprosium
material {[}\{(CuL)$_{2}$Dy\}\{Mo(CN)$_{8}$\}{]} which can be represented
experimentally through an Ising-Heisenberg chain as a {[}DyCuMoCu{]} infinite
chain\cite{vanden10,bellucci}, and the single chain magnet {[}\{(H$_{2}$O)Fe(L)\}\{Nb(CN)$_{8}$\}\{Fe(L)\}{]}
\cite{sahoo12} was well described within the framework of the Ising-Heisenberg
chains.

Here, we consider the alternating Ising and Heisenberg spin chain
under arbitrarily oriented magnetic field, which nicely describes
the 3d-4f bimetallic polymeric compound Dy(NO$_{3}$)(DMSO)$_{2}$Cu(opba)(DMSO)$_{2}$\cite{hagiwara-strecka},
which provides an interesting experimental realization of the ferrimagnetic
chain composed of two different but regularly alternating spin-1/2
magnetic ions Dy$^{3+}$ and Cu$^{2+}$ that are reasonably approximated
by the notion of Ising and Heisenberg spins, respectively.

\paragraph{1. The alternating Ising and Heisenberg spin chain in an arbitrarily
oriented magnetic field.}

Let us consider the following Hamiltonian under an arbitrarily oriented
magnetic field,
\begin{equation}
\mathcal{H}=-\sum_{i=1}^{N}\left((J\sigma_{i}^{z}+\frac{h_{1}}{2})(s_{i}+s_{i+1})+\boldsymbol{h}.\boldsymbol{\sigma}_{i}\right),
\end{equation}
where $\boldsymbol{h}.\boldsymbol{\sigma}_{i}=h_{z}\sigma_{i}^{z}+h_{x}\sigma_{i}^{x}+h_{y}\sigma_{i}^{y}$,
with $h_{x}=g_{x}h\sin(\phi)\cos(\theta)$, $h_{y}=g_{y}h\sin(\phi)\sin(\theta)$,
$h_{z}=g_{z}h\cos(\phi)$ and $h_{1}=g_{1,z}h\cos(\phi)$, assuming
$\theta\in[0,2\pi]$ and $\phi\in[0,\pi]$, and $\sigma^{\alpha}$
is the Pauli matrix for the Heisenberg spin, while $s$ corresponds
to the Ising spin ($s=\pm1)$. The exchange coupling parameter between
$s$ and $\sigma^{z}$ spins is given by $J$.

Writing in  matrix form, the Hamiltonian for the unit cell becomes
\begin{equation}
\mathcal{H}_{i,i+1}(\mu)=\left(\begin{array}{cc}
-(J\mu+h_{z})-\frac{h_{1}\mu}{2} & -h_{x}+\hat{\imath}h_{y}\\
-h_{x}-\hat{\imath}h_{y} & (J\mu+h_{z})-\frac{h_{1}\mu}{2}
\end{array}\right),\label{eq:H-mat}
\end{equation}
 with $\mu=s_{i}+s_{i+1}$.

To use the decoration transformation\cite{strecka pla,JPA-11},
we need to diagonalize the Hamiltonian \eqref{eq:H-mat}, whose eigenvalues
are given by
\begin{equation}
\mathcal{E}_{i,i+1}(\mu)=-\frac{h_{1}}{2}\mu\pm A(\mu),
\end{equation}
where $A(\mu)=\sqrt{(J\mu+h_{z})^{2}+h_{x}^{2}+h_{y}^{2}}$. 

This model can be solved exactly through a decoration transformation\cite{strecka pla,JPA-11}
and transfer matrix approach\cite{baxter-book}.

To compute all thermodynamic quantities, we need to calculate the Boltzmann
factor given by

\begin{alignat}{1}
w(\mu) & =\text{tr}_{\sigma}\left({\rm e}^{-\beta\mathcal{H}_{i,i+1}(\mu)}\right)=2{\rm e}^{\beta\mu\frac{h_{1}}{2}}\cosh(-\beta A(\mu)).
\end{alignat}
Once  the Boltzmann factor is known, the transfer matrix can be expressed
as follow
\begin{equation}
\boldsymbol{W}=\left(\begin{array}{cc}
w(2) & w(0)\\
w(0) & w(-2)
\end{array}\right).
\end{equation}
Furthermore, the eigenvalues of the transfer matrix can be expressed
by
\begin{equation}
\lambda_{\pm}=\frac{w(2)+w(-2)}{2}\pm\frac{B}{2},
\end{equation}
with $B=\sqrt{(w(2)-w(-2))^{2}+4w(0)^{2}}$.

The corresponding non singular matrix that diagonalize the transfer
matrix $\boldsymbol{T}$, becomes 
\begin{equation}
\boldsymbol{V}=\left(\begin{array}{cc}
b_{+} & b_{-}\\
1 & 1
\end{array}\right)\;\text{and}\;\boldsymbol{V}^{-1}=\tfrac{1}{b_{-}-b_{+}}\left(\begin{array}{cc}
1 & -b_{-}\\
-1 & b_{+}
\end{array}\right),
\end{equation}
with $b_{\pm}=\frac{w(2)-w(-2)\pm B}{2w(0)}$.

\paragraph{2. The correlation function.}

To perform the expectation value and correlation function,
we need to perform the following quantity
\begin{equation}
w_{\alpha}(\mu)={\rm tr}_{\sigma}\left(\sigma_{i}^{\alpha}{\rm e}^{-\beta\mathcal{H}_{i,i+1}(\mu)}\right).
\end{equation}
In fact, the trace of these quantities can be expressed easily as
the derivative of the Boltzmann factor, $w_{\alpha}(\mu)=\frac{\partial w(\mu)}{\beta\partial h_{\alpha}},$
thus, each derivatives components becomes $w_{x}(\mu)=h_{x}\bar{w}(\mu)$,
$w_{y}(\mu)=h_{y}\bar{w}(\mu)$ and $ $$w_{z}(\mu)=(J\mu+h_{z})\bar{w}(\mu),$
where
\begin{equation}
\bar{w}(\mu)=\frac{2{\rm e}^{\beta\mu\frac{h_{1}}{2}}\sinh(-\beta A(\mu))}{A(\mu)}.
\end{equation}
Thereafter, writing in  matrix form, we have
\begin{equation}
\boldsymbol{W}_{\alpha}=\left(\begin{array}{cc}
w_{\alpha}(2) & w_{\alpha}(0)\\
w_{\alpha}(0) & w_{\alpha}(-2)
\end{array}\right).
\end{equation}
The expectation value of $\sigma^{\alpha}$ is denoted by $\langle\sigma^{\alpha}\rangle$,
thus in thermodynamic limit ($N\rightarrow\infty$) one can write
as follow
\begin{equation}
\langle\sigma^{\alpha}\rangle=\frac{1}{\lambda_{+}}{\rm tr}\left[\boldsymbol{V}^{-1}\boldsymbol{W}_{\alpha}\boldsymbol{V}\left(\begin{array}{cc}
1 & 0\\
0 & 0
\end{array}\right)\right],
\end{equation}
defining $\boldsymbol{V}^{-1}\boldsymbol{W}_{\alpha}\boldsymbol{V}=\frac{1}{b_{+}-b_{-}}\widetilde{\boldsymbol{W}}_{\alpha}$,
with $\widetilde{\boldsymbol{W}}_{\alpha}$ given by $\widetilde{\boldsymbol{W}}_{\alpha}=\left(\begin{array}{cc}
\nu_{11}^{\alpha} & \nu_{12}^{\alpha}\\
\nu_{21}^{\alpha} & \nu_{22}^{\alpha}
\end{array}\right)$, whose elements are 
\begin{alignat}{1}
\nu_{11}^{\alpha}= & b_{+}w_{\alpha}(2)+2w_{\alpha}(0)-b_{-}w_{\alpha}(-2),\\
\nu_{12}^{\alpha}= & b_{-}w_{\alpha}(2)+(1-b_{-}^{2})w_{\alpha}(0)-b_{-}w_{\alpha}(-2),\\
\nu_{21}^{\alpha}= & -b_{+}w_{\alpha}(2)-(1-b_{+}^{2})w_{\alpha}(0)+b_{+}w_{\alpha}(-2),\\
\nu_{22}^{\alpha}= & -b_{-}w_{\alpha}(2)-2w_{\alpha}(0)+b_{+}w_{\alpha}(-2).
\end{alignat}
After taking the trace and assuming $b_{+}-b_{-}=\frac{B}{w(0)}$,
the correlation function goes to
\begin{equation}
\langle\sigma^{\alpha}\rangle=\frac{w(0)}{B\lambda_{+}}\left(b_{+}w_{\alpha}(2)+2w_{\alpha}(0)-b_{-}w_{\alpha}(-2)\right).
\end{equation}
Consequently, we define the following quantities
\begin{alignat}{1}
m_{0} & =\frac{w(0)}{B\lambda_{+}}\left(b_{+}\bar{w}(2)+2\bar{w}(0)-b_{-}\bar{w}(-2)\right),\\
m_{1} & =\frac{w(0)}{B\lambda_{+}}\left(b_{+}\bar{w}(2)+b_{-}\bar{w}(-2)\right),
\end{alignat}
thus, the expectation value of $\sigma^{\alpha}$ become

\begin{equation}
\langle\sigma^{x}\rangle=h_{x}m_{0},\;\langle\sigma^{y}\rangle=h_{y}m_{0},\;\langle\sigma^{z}\rangle=h_{z}m_{0}+2Jm_{1}.
\end{equation}
It is worth noting that, the expectation value $\langle\sigma^{\alpha}\rangle$
can be expressed only in terms of the Boltzmann factor $w(\mu)$ and its
derivative $w_{\alpha}(\mu)$, since, $b_{\pm}$ and $B$  were also
defined in terms of the Boltzmann factor.

In a similar way, we can develop the correlation function in the thermodynamic
limit,
\begin{equation}
\langle\sigma_{i}^{\alpha}\sigma_{i+r}^{\alpha^{\prime}}\rangle=\frac{w(0)^{2}}{B^{2}\lambda_{+}^{2}}{\rm tr}\left[\widetilde{\boldsymbol{W}}_{\alpha}\left(\begin{array}{cc}
1 & 0\\
0 & q^{r-1}
\end{array}\right)\widetilde{\boldsymbol{W}}_{\alpha^{\prime}}\left(\begin{array}{cc}
1 & 0\\
0 & 0
\end{array}\right)\right],
\end{equation}
where $q$ is the transfer matrix eigenvalues ratio $q=\frac{\lambda_{-}}{\lambda_{+}}$.
After some algebraic manipulation, the correlation function becomes
\begin{alignat}{1}
\langle\sigma_{i}^{\alpha}\sigma_{i+r}^{\alpha^{\prime}}\rangle= & \langle\sigma^{\alpha}\rangle\langle\sigma^{\alpha^{\prime}}\rangle+\left(\tfrac{w_{\alpha}(2)-w_{\alpha}(-2)-(b_{-}+b_{+})w_{\alpha}(0)}{(b_{+}-b_{-})\lambda_{+}}\right)\times\nonumber \\
 & \left(\tfrac{w_{\alpha^{\prime}}(2)-w_{\alpha^{\prime}}(-2)-(b_{-}+b_{+})w_{\alpha^{\prime}}(0)}{(b_{+}-b_{-})\lambda_{+}}\right)q^{r-1}.
\end{alignat}
For a particular case when $\alpha^{\prime}=\alpha$, it is worth
noting that all correlation function of the form $\langle\sigma_{i}^{\alpha}\sigma_{i+r}^{\alpha}\rangle$
are always positive. Using a convenient notation, 
\begin{alignat}{1}
p_{0} & =\frac{w(0)}{B\lambda_{+}}\left(\bar{w}(2)+(b_{+}+b_{-})\bar{w}(0)-\bar{w}(-2)\right),\\
p_{1} & =\frac{w(0)}{B\lambda_{+}}\left(\bar{w}(2)+\bar{w}(-2)\right).
\end{alignat}
we will describe the correlation function as follow,
\begin{alignat}{1}
\langle\sigma_{i}^{x}\sigma_{i+r}^{x}\rangle= & \langle\sigma^{x}\rangle^{2}+h_{x}^{2}p_{0}^{2}q^{r-1},\\
\langle\sigma_{i}^{y}\sigma_{i+r}^{y}\rangle= & \langle\sigma^{y}\rangle^{2}+h_{y}^{2}p_{0}^{2}q^{r-1},\\
\langle\sigma_{i}^{x}\sigma_{i+r}^{y}\rangle= & \langle\sigma^{x}\rangle\langle\sigma^{y}\rangle+h_{x}h_{y}p_{0}^{2}q^{r-1},\\
\langle\sigma_{i}^{z}\sigma_{i+r}^{z}\rangle= & \langle\sigma^{z}\rangle^{2}+\left(p_{0}h_{z}+2Jp_{1}\right)^{2}q^{r-1}.
\end{alignat}

Once again, all correlation for Heisenberg spins can be expressed
only in terms of the Boltzmann factor and its derivative.

\paragraph{3. Thermal entanglement.}

Despite the system being untangled at zero temperature, it is of great relevance to
 discuss the entanglement at a non-zero temperature. Thus, we consider
the quantum entanglement between a pair of Heisenberg spins in an
arbitrarily oriented magnetic field.

As a measure of entanglement for two arbitrary mixed states of two
qubits, we use the quantity called concurrence\cite{wooters}, which
was defined in terms of reduced density matrix $\rho$ of two mixed
states,
\begin{equation}
\mathcal{C}(\rho)=\max\{{0,2\Lambda_{{\rm max}}-\text{{tr}}\sqrt{R}}\},
\end{equation}
assuming $R=\rho\sigma^{y}\otimes\sigma^{y}\rho^{*}\sigma^{y}\otimes\sigma^{y},$
where $\Lambda_{{\rm max}}$ is the largest eigenvalue of the matrix
$\sqrt{R}$, and $\rho^{*}$ represents the complex conjugate of matrix
$\rho$, with $\sigma^{y}$ being the Pauli matrix. 

For the case of an infinite chain, the reduced density operator elements\cite{bukman}
could be expressed in terms of the correlation function between two
entangled particles\cite{amico}, thus we have
\begin{alignat}{1}
\mathcal{C}_{r}(\theta,\phi)= & \frac{1}{2}\max\left\{ P_{r}^{xy}-\left|1-\langle\sigma_{i}^{z}\sigma_{i+r}^{z}\rangle^{2}\right|,0\right\} 
\end{alignat}
where $P_{r}^{xy}=\sqrt{\left(\langle\sigma_{i}^{x}\sigma_{i+r}^{x}\rangle-\langle\sigma_{i}^{y}\sigma_{i+r}^{y}\rangle\right)^{2}+4\langle\sigma_{i}^{x}\sigma_{i+r}^{y}\rangle^{2}}$. 

For the particular case when $g_{x}=g_{y}$, the concurrence becomes
independent of  $\theta$, and reduces to
\begin{equation}
\mathcal{C}_{r}(\phi)=\frac{1}{2}\max\left\{ 0,2\langle\sigma_{i}^{x}\sigma_{i+r}^{x}\rangle+\langle\sigma_{i}^{z}\sigma_{i+r}^{z}\rangle-1\right\} .\label{eq:Cnr-1}
\end{equation}

Using this result, we will illustrate the entangled region and how
the concurrence behave. Alternatively, we can also obtain an equivalent result using the approach described in Ref \cite{spra}.
\begin{figure}
\includegraphics[scale=0.12]{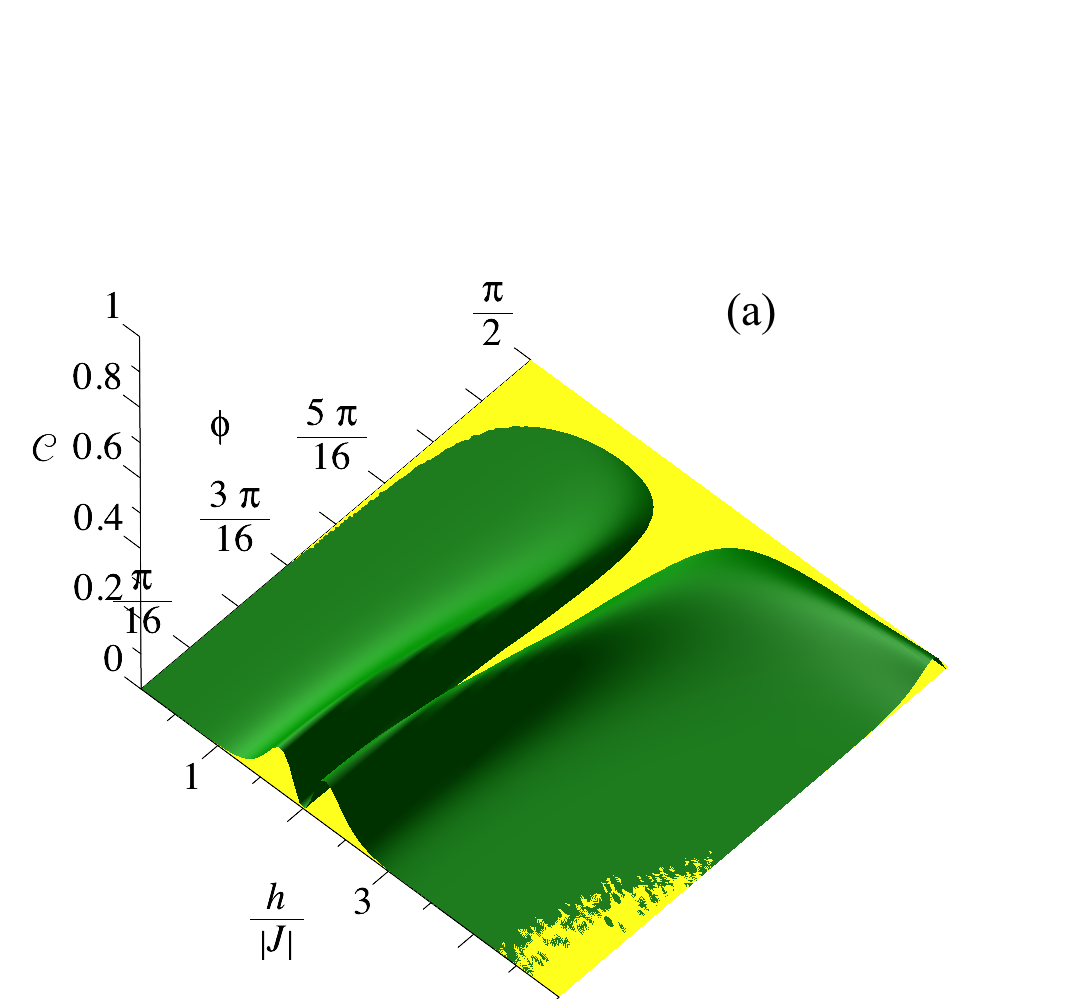}\includegraphics[scale=0.12]{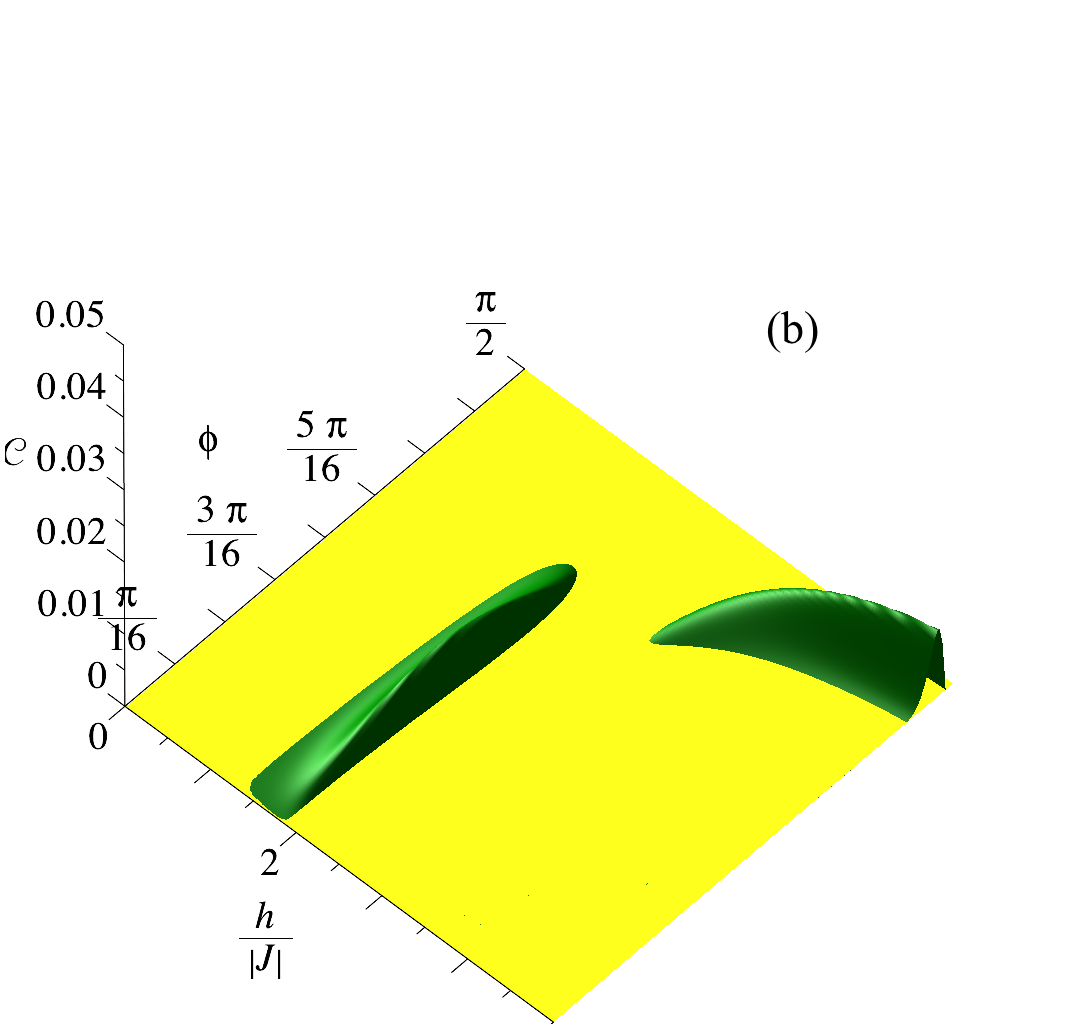}

\protect\caption{\label{fig:h-phi}(Color online) The arising of pairwise thermal entanglement
(green region). (a) The concurrence $\mathcal{C}$ between the nearest
Heisenberg spin $r=1$, as a function of the magnetic field $h/|J|$ and the
polar angle $\phi$. (b) The concurrence $\mathcal{C}$ between the next-nearest
Heisenberg spin $r=2$, as a function of the magnetic field $h/|J|$ and the 
polar angle $\phi$. }
\end{figure}

In figure \ref{fig:h-phi}, we illustrate the arising of pairwise
thermal entanglement [green (gray) region], while the yellow region corresponds
to the untangled region. By the use of concurrence, we display the intensity
of entanglement, assuming the following gyromagnetic factors: $g_{x}=1.5$,
$g_{z}=2.0$, and $g_{1z}=1.0$. We also choose the coupling parameter
$J=-1.0$, while the temperature is assumed to be $T/|J|=0.3$. In figure
\ref{fig:h-phi}(a), we illustrate the concurrence $\mathcal{C}$
as a function of magnetic field and polar angle $\phi$ for the nearest
neighbor of the Heisenberg spin ($r=1$). But actually it means the second
nearest neighbor, because there is one Ising spin between two Heisenberg
spins. There is an entangled region for a relatively low magnetic field,
$h/|J|\lesssim2$, while for $h/|J|\gtrsim2$ there is another entangled region.
It is worth  mentioning the dependence of the polar angle $\phi$. When a
pure transverse magnetic field is applied to an alternating spin chain,
this entanglement vanishes at $\phi=\frac{\pi}{2}$. Despite the fact that the transverse
magnetic field ($\phi=\frac{\pi}{2}$) never generates concurrence
for a high magnetic field,  a relatively intense concurrence appears
in the limit of $\phi\rightarrow\frac{\pi}{2}$, while for $\phi<\frac{\pi}{2}$
and a high magnetic field the concurrence vanishes. Similarly, figure
\ref{fig:h-phi}(b) illustrates the concurrence $\mathcal{C}$ for
the case of $r=2$ (next-nearest neighbor between Heisenberg spins). 
It should be noted that between this pair of Heisenberg spins, there are two Ising
spins. So it is still possible to observe a weak concurrence around
$\mathcal{C}\thickapprox0.01$, however the entangled (green) region
shrunk significantly. The entanglement for longer pair spins still
appears for $r=3$ and $4$, with maximum concurrence $\mathcal{C}\thickapprox0.001$,
and for a tiny specific region, which is not illustrated here due to the  irrelevant
amount of concurrence.

\begin{figure}
\includegraphics[scale=0.12]{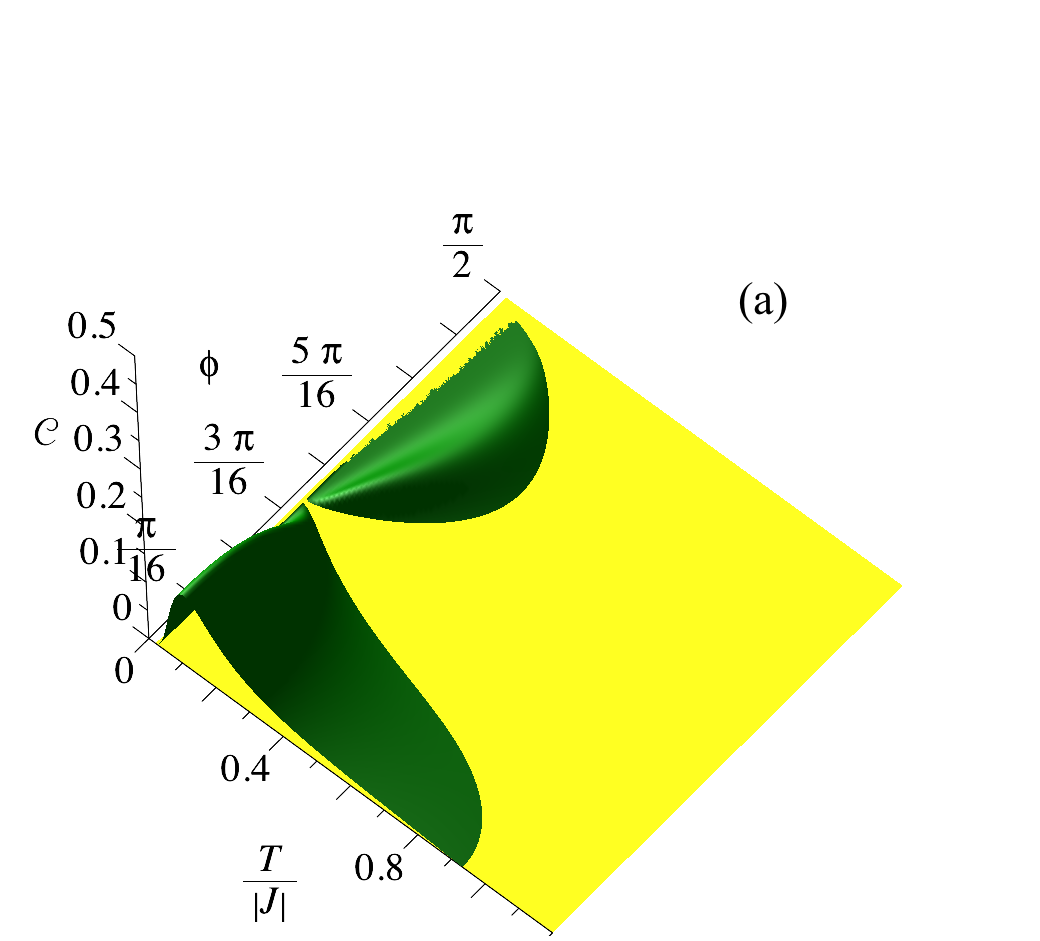}\includegraphics[scale=0.125]{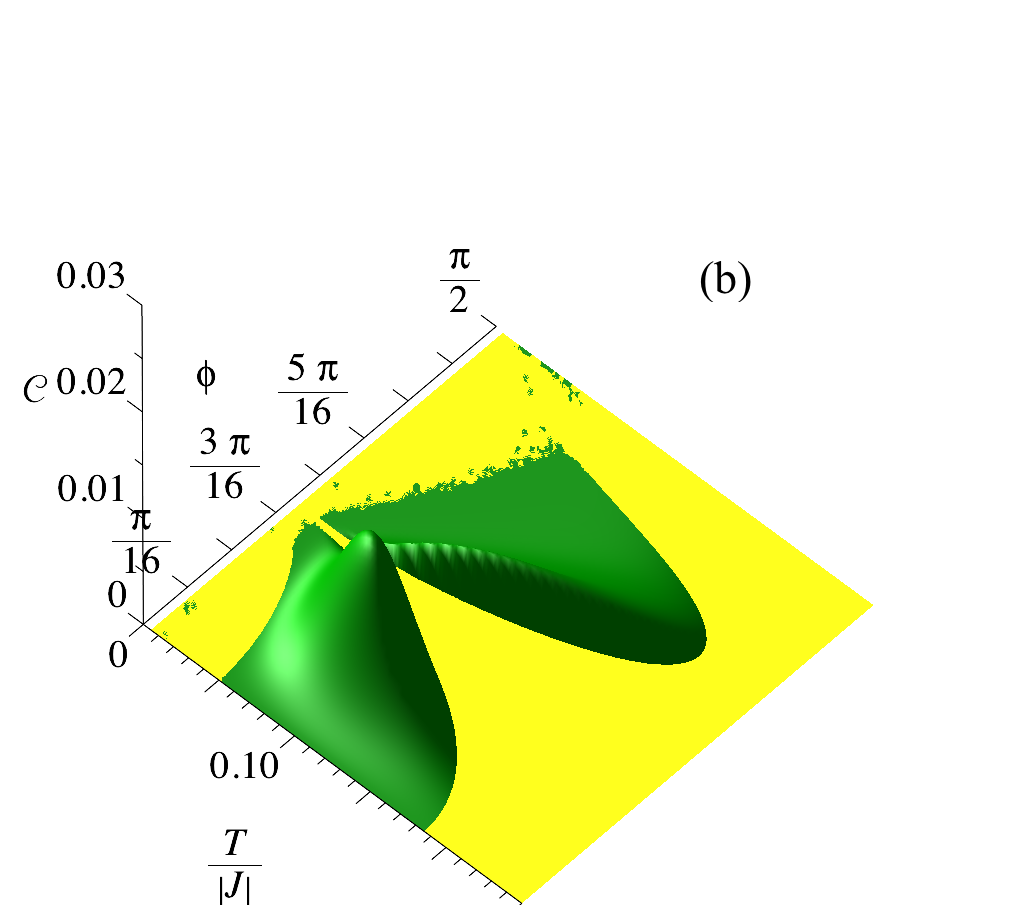}\protect\caption{\label{fig:T-phi}(Color online) The arising of pairwise thermal entanglement
(green region). (a) The concurrence $\mathcal{C}$ between nearest the 
Heisenberg spin $r=1$, as a function of the temperature $T/|J|$ and the polar
angle $\phi$. (b) The concurrence $\mathcal{C}$ between the next-nearest
Heisenberg spin $r=2$, as a function of the temperature $T/|J|$ and the  polar
angle $\phi$. }
\end{figure}

In figure \ref{fig:T-phi}(a) we illustrate the concurrence $\mathcal{C}$
as a function of temperature and polar angle $\phi$ for nearest neighbor
of Heisenberg spin ($r=1$), for a fixed magnetic field $h/|J|=2.1$.
There is a stronger entangled region for $0<\phi\lesssim\frac{3\pi}{16}$,
while for $\frac{3\pi}{16}\lesssim\phi<\frac{\pi}{2}$, a weaker entangled
region is observed, it is worth to notice the temperature dependence
of concurrence disappear at the threshold temperature. Similarly,
figure \ref{fig:T-phi}(b) also illustrates the concurrence $\mathcal{C}$
for the case of $r=2$.

\paragraph{4. Concurrence for powder samples.}

Synthesized real materials are usually  known as the single molecule
magnets chains\cite{vanden10,sahoo12,hagiwara-strecka}, so it is
natural to define the concurrence for the powder samples as an average
of concurrence for powder samples. In what follows, we will discuss
our theoretical results with measurements performed on powder samples
of the crystalline compounds.
Let $\theta$ and $\phi$ be azimuthal and polar angles of the magnetic
field vector with respect to a molecular reference frame; thus the
powder sample concurrence can be defined as a function of $\theta$
and $\phi$ ,
\begin{equation}
\mathcal{\bar{C}}_{r}=\frac{1}{4\pi}\int_{0}^{\pi}\int_{0}^{2\pi}\mathcal{C}_{r}(\theta,\phi)\sin(\phi){\rm d}\theta{\rm d}\phi.\label{eq:av-C}
\end{equation}
For the case of $g_{x}=g_{y}$, the concurrence is independent of
the angle $\theta$; thus we only need to integrate over the polar
angle $\phi$. Hereafter, we observe the concurrence for the powder
sample as a function of magnetic field and temperature.

\begin{figure}
\includegraphics[scale=0.22]{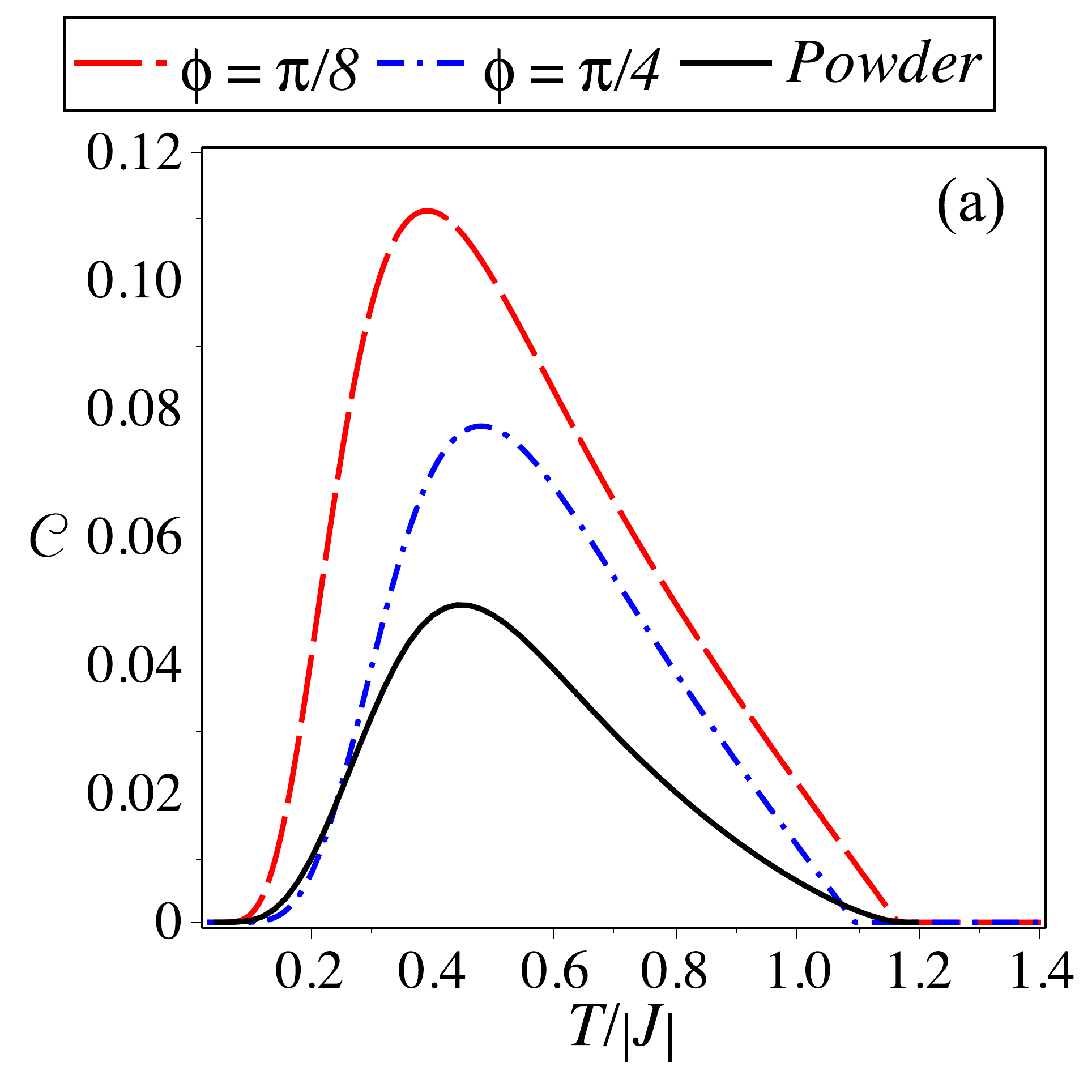}\includegraphics[scale=0.22]{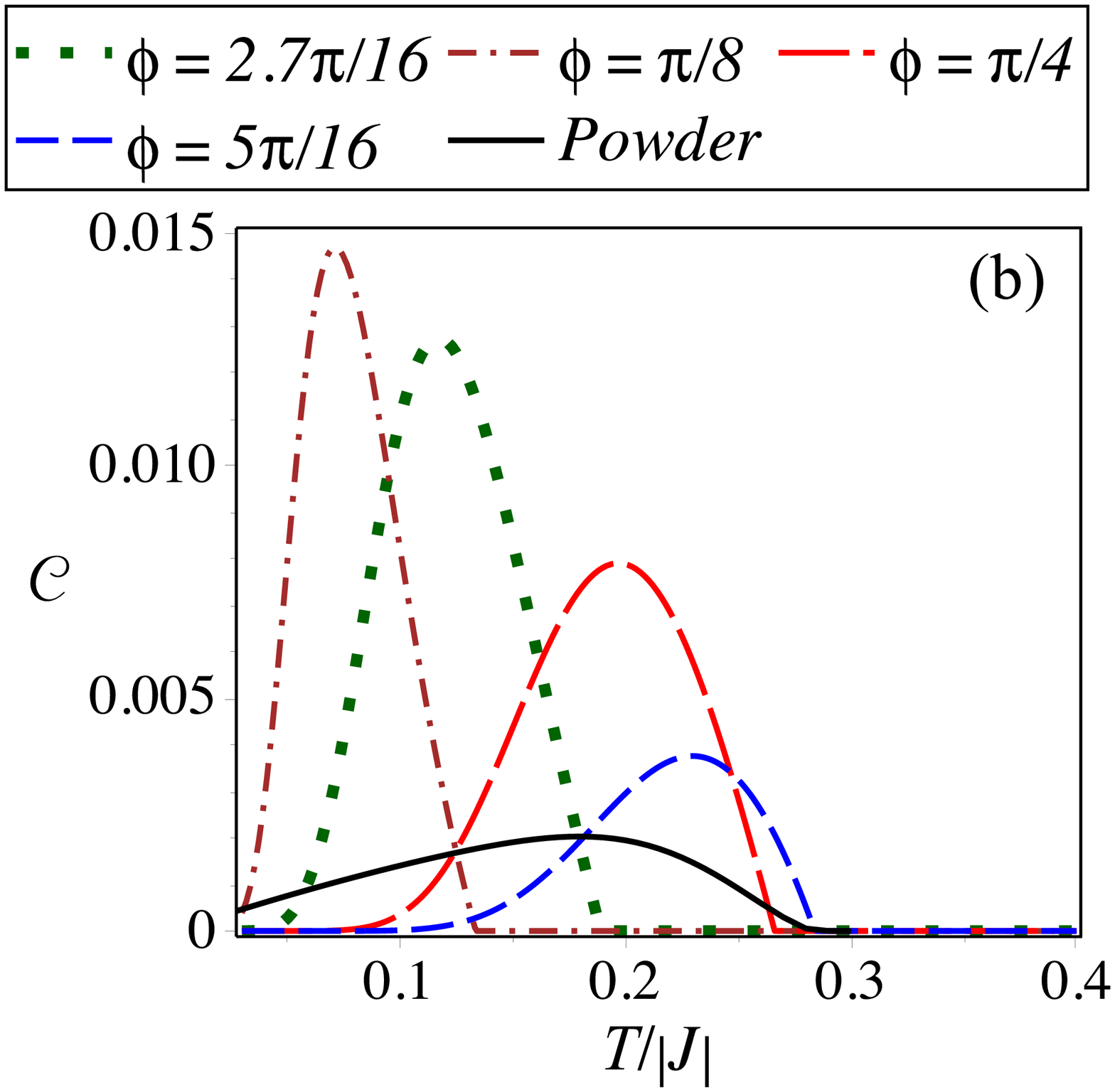}\protect\caption{\label{fig:TC}(Color online) Concurrence as a function of temperature
for a fixed value of magnetic field. (a) Nearest-neighbor concurrence
$r=1$. (b) Next-nearest-neighbor concurrence $r=2$.}
\end{figure}

In figure \ref{fig:TC}(a) is illustrated the nearest-neighbor concurrence
($r=1$), assuming fixed values $h/|J|=1.7$ and $J=-1$. The black solid
line corresponds to the powder sample concurrence given in eq.\eqref{eq:av-C},
while the dashed line corresponds to the concurrence for a fixed
value of $\phi=\pi/8$, with the maximum concurrence $\mathcal{C}\approx0.11$
at around $T/|J|\approx0.4$. For other angles, the concurrence vanishes
quickly, such as for the dashed-dotted line representing the concurrence
for $\phi=\pi/4$, whose maximum occurs at $\mathcal{C}\approx0.08$
and for $T/|J|\approx0.5$. In contrast,  figure \ref{fig:TC}(b) corresponds
to the next-nearest-neighbor (NNN) pairwise concurrence ($r=2$),
taking into account fixed values $h/|J|=2.0$ and $J=-1$. The black solid line
represents to the powder sample concurrence, the dotted line
corresponds to the concurrence for a fixed value of $\phi=2.7\pi/16$,
the dashed-dotted line represents the concurrence for $\phi=\pi/8$,
 the long-dashed line corresponds to $\phi=\pi/4$, and finally
the dashed line represents the concurrence for $\phi=5\pi/16$.

\begin{figure}
\includegraphics[scale=0.22]{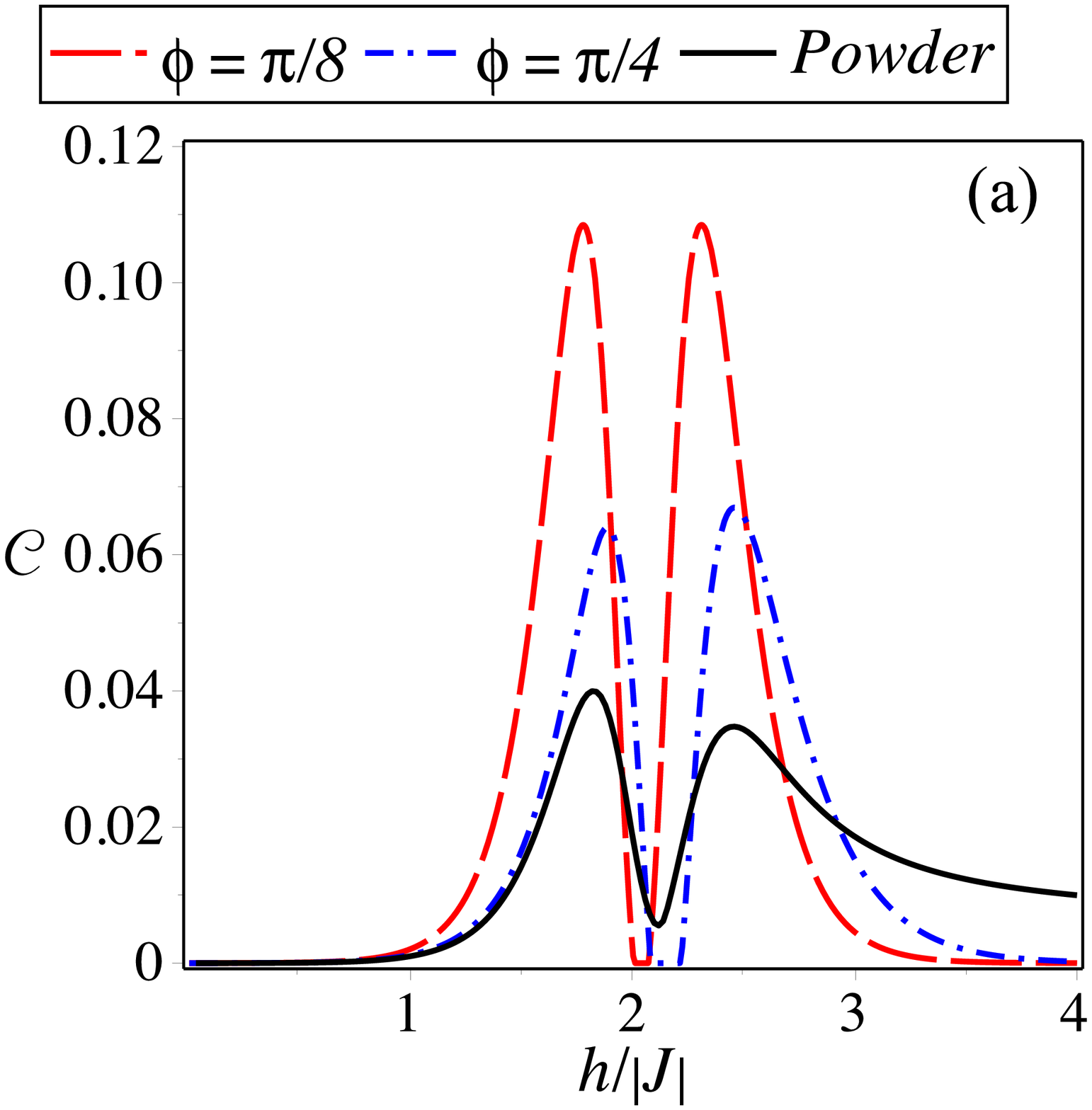}\includegraphics[scale=0.22]{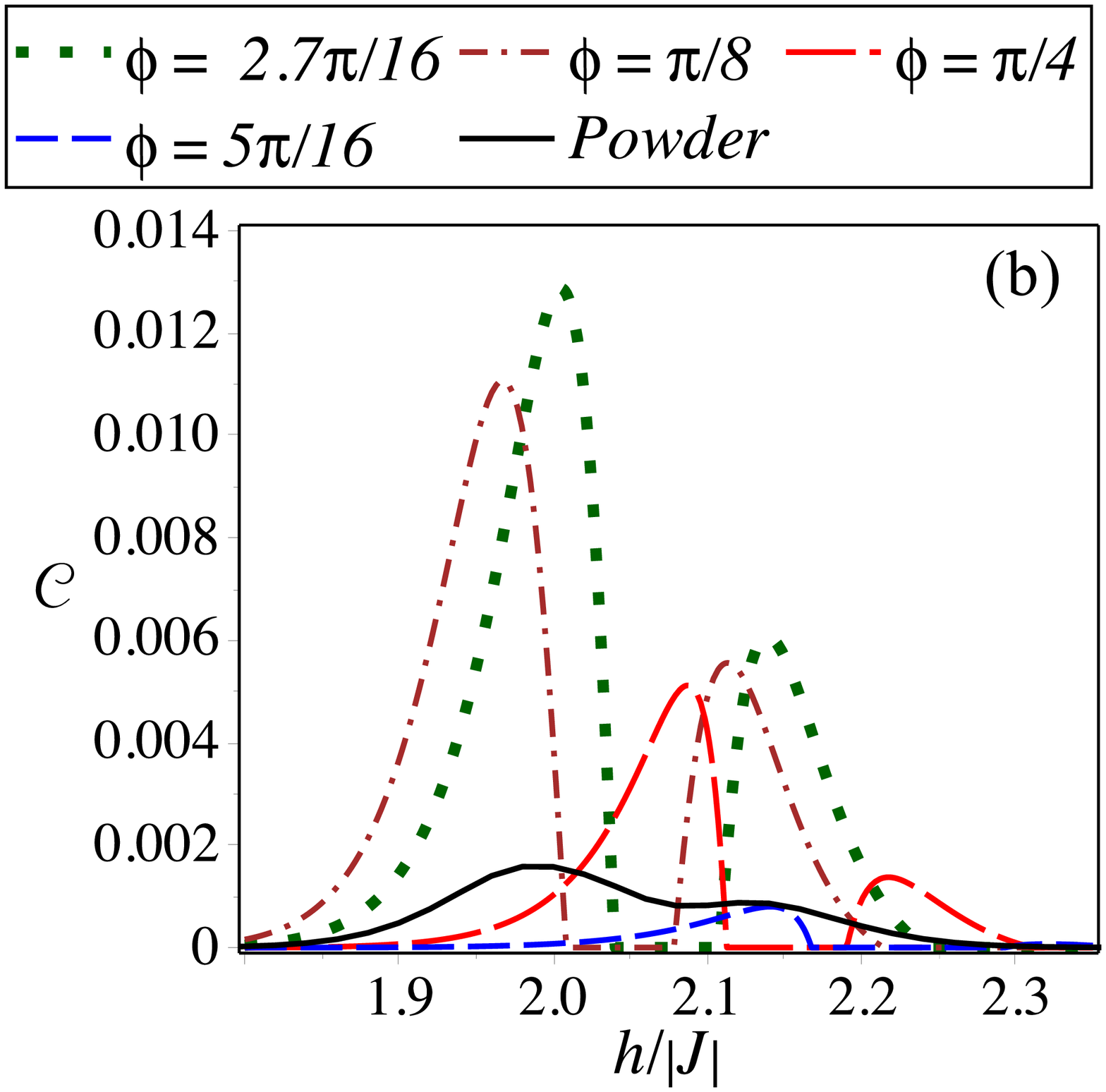}\protect\caption{\label{fig:hC}(Color online) Concurrence as a function of magnetic
field for a fixed temperature. (a) Nearest-neighbor concurrence $r=1$.
(b) Next-nearest-neighbor concurrence $r=2$.}
\end{figure}

Furthermore, in figure \ref{fig:hC} is illustrated the concurrence
as a function of magnetic field for a fixed temperature and $J=-1$.
In fig.\ref{fig:hC}(a), we illustrate the concurrence for nearest-neighbor
($r=1$) and fixed temperature $T/|J|=0.3$. The black solid line corresponds
to the powder sample concurrence given in eq.\eqref{eq:av-C}. While
the dashed line corresponds to the concurrence for a fixed value of
$\phi=\pi/8$, with maximum concurrence $\mathcal{C}\approx0.11$
at around $h/|J|\approx1.7$ and $h/|J|\approx2.5$. For other angles, concurrence
vanishes quickly, such as for the dashed-dotted line representing the
concurrence for $\phi=\pi/4$, with a maximum at around $\mathcal{C}\approx0.06$
and for an external magnetic field $h/|J|\approx0.75$ and $h/|J|\approx2.6$.
The fig.\ref{fig:hC}(b) corresponds to NNN pairwise concurrence
$r=2$ and assuming $T/|J|=0.115$. The black solid line corresponds to
powder sample concurrence, the dashed line corresponds to the concurrence
for a fixed value of $\phi=5\pi/16$, the dashed-dotted line represents
the concurrence for $\phi=\pi/8$, the dotted line is associated with $\phi=2.7\pi/16$,
and finally the long-dashed line corresponds to the concurrence for $\phi=\pi/4$.

\paragraph{5. Conclusion.}

The considered spin-chain model provides as insight into the  3d-4f bimetallic
polymeric compound Dy(NO$_{3}$)(DMSO)$_{2}$Cu(opba)(DMSO)$_{2}$\cite{hagiwara-strecka},
which provides an interesting experimental realization of the ferrimagnetic
chain composed of two different but regularly alternating spin-$\frac{1}{2}$
magnetic ions Dy$^{3+}$ and Cu$^{2+}$ that are nearly well represented
by Ising and Heisenberg spins. To solve the one-dimensional Ising
model with alternating Ising and Heisenberg spins, one can map onto
the classical Ising model. With regard to real material studied by Han
et al.\cite{Han-strecka}, we have assumed the following factors:
$g_{x}=2.0$, $g_{z}=2.0$ and $g_{1z}=20$, and the coupling parameter
$J=-26$ given in reference\cite{Han-strecka,hagiwara-strecka}. Therefore,
the theoretical prediction for the concurrence would be possible at
$T=2.5$, but for an ultrahigh magnetic field above  $h=50$, so
we believe this should be difficult to measure. Nevertheless, one
thing we can claim is that for  strong factors $g_{x}=g_{y}$ the concurrence
arises yet for a lower magnetic field, which was discussed in this paper. 

Typically, two particles (spins) are maximally entangled at zero temperature,
and such a  phenomenon could vanish at the
threshold temperature. However, at finite temperature,  pairwise
entanglement emerges surprisingly, for an arbitrarily oriented magnetic
field. This effect, is purely due to the magnetic field and the temperature
dependence, i.e. as soon as the temperature increases arises a small amount
of concurrence between nearest-neighbor spins taking its maximum at
around 0.1.

This work was supported by the Brazilian agencies: FAPEMIG and CNPq, M. Rojas also thanks  CAPES for fully financial support.

\end{document}